\documentclass[twocolumn,apj]{aastex63}
\usepackage{apjfonts}
\usepackage{float}
\usepackage{longtable}
\usepackage[titletoc, title]{appendix}
\usepackage{amsmath}
\usepackage{array}
\usepackage{tabularx}
\usepackage{lineno}
\usepackage{textgreek}
\hypersetup{linkcolor=red,citecolor=blue,filecolor=red,urlcolor=blue}

\graphicspath{{./}{figures/}}

\newcommand{\rsi}{Rev. Sci. Inst.} 

\shorttitle{\ion{Fe}{17} DR Cross Sections}
\shortauthors{Grell et al. 2024}

\begin{document}

\title{\large\textsc{Laboratory Benchmark of $n\geq4$ Dielectronic Recombination Satellites of~\ion{Fe}{17}}}

\author[0000-0003-3363-9786]{Gabriel J. Grell}
\affiliation{University of Maryland College Park, College Park, MD 20742, USA; \href{mailto:ggrell@umd.edu}{ggrell@umd.edu}}
\affil{NASA/Goddard Space Flight Center, 8800 Greenbelt Road, Greenbelt, MD 20771, USA}%
\affiliation{CRESST II, Greenbelt, MD 20771, USA}

\author[0000-0002-3331-7595]{Maurice A. Leutenegger}
\affil{NASA/Goddard Space Flight Center, 8800 Greenbelt Road, Greenbelt, MD 20771, USA}%

\author[0000-0002-5257-6728]{Pedro Amaro}
\affil{Laboratory of Instrumentation, Biomedical Engineering and Radiation Physics (LIBPhys-UNL), Department of Physics, NOVA School of Science and Technology, NOVA University Lisbon, 2829-516 Caparica, Portugal}

\author[0000-0002-2937-8037]{Jos\'e R. {Crespo L\'opez-Urrutia}}
\affil{Max-Planck-Institut f\"ur Kernphysik, Saupfercheckweg 1, 69117 Heidelberg, Germany;\href{mailto:chintan.shah@mpi-hd.mpg.de}{chintan.shah@mpi-hd.mpg.de}}

\author[0000-0002-6484-3803]{Chintan Shah}
\affil{NASA/Goddard Space Flight Center, 8800 Greenbelt Road, Greenbelt, MD 20771, USA}%
\affil{Max-Planck-Institut f\"ur Kernphysik, Saupfercheckweg 1, 69117 Heidelberg, Germany;\href{mailto:chintan.shah@mpi-hd.mpg.de}{chintan.shah@mpi-hd.mpg.de}}
\affil{Department of Physics and Astronomy, Johns Hopkins University, Baltimore, MD 21218, USA}

\begin{abstract}

We calculated cross sections for the dielectronic recombination (DR) satellite lines of \ion{Fe}{17} and benchmarked our predictions with experimental cross sections of \ion{Fe}{17} resonances that were mono-energetically excited in an electron beam ion trap. We extend the benchmark to all resolved DR and direct electron-impact excitation (DE) channels in the experimental dataset, specifically the $n\geq4$ DR resonances of \ion{Fe}{17}, complementing earlier investigations of $n=3$ channels. Our predictions overestimate by 20-25$\%$ the DR and DE absolute cross sections for the higher $n$ complexes when using the same methods as in previous works. However, we achieve agreement within $\sim$10$\%$ of the experimental results by an approach in which we "forward fold" the predicted cross sections with the spread of the electron-beam energy and the photon-energy resolution of our experiment. We then calculated rate coefficients from the experimental and theoretical cross sections, finding departures of $10-20\%$ from the rates found in the OPEN-ADAS atomic database.

\end{abstract}


\section{Introduction}
\label{sec:intro}

Some of the strongest features in the X-ray spectra of many collisional plasma sources, including coronal and massive stars, galaxy clusters, the interstellar medium, and X-ray binaries \citep{par1973, smith1985, schmelz92, wmd1994, Phillips96, behar2001, MLF01, doron2002, xpb2002, fgu2003, pfk2003, werner2009, pradhan2011, beiersdorfer2018, gu2020} are due to the Fe-L complex. It encompasses radiative transitions from $n=2$ states of Na-like (\ion{Fe}{16}) to Li-like (\ion{Fe}{23}) Fe ions, primarily excited by electronic impact, recombination, and ionization \citep{Gu2019}. Within this complex, and due to its closed-shell configuration and correspondingly high ionization potential \citep{smith1985}, neon-like \ion{Fe}{17} displays  some of the brightest spectral signatures of any highly charged ion seen in hot astrophysical plasmas. Their spectra at temperatures of a few MK are dominated by the L-shell transitions of \ion{Fe}{17} ions in the $15-18$ $\mbox{\AA}$ range, and specifically the $3d\mbox{ -- }2p$ and $3s\mbox{ -- }2p$ transitions \citep{par1973,chd2000,behar2001,xpb2002,PK2003}. 

These transitions also provide very useful diagnostics of the physical conditions in such plasmas, including electron temperature as well as density, velocity turbulence, and X-ray opacity \citep{behar2001, mrd2001, PK2003, kem2014, beiersdorfer2018, grell2021}. Decades of laboratory measurements have yielded accurate wavelengths, cross sections, and intensity ratios of those transitions \citep{B98,bbc2001diag, bbc2001sys, bei2002, bbg2004, brown2006, glt2011, bbl2017, Shah19, shah2024arxiv}. However, their diagnostic utility is hampered by the clear discrepancies between observations, laboratory measurements, and theoretical calculations of their relative line intensities that were found. One of the key line formation mechanisms for \ion{Fe}{17} in hot plasmas, direct electron-impact excitation (DE), has exhibited $10-20\%$ model-data disparities for the $3d\mbox{ -- }2p$ cross sections in numerous studies over several decades \citep{B98, lkt2000,bbc2001diag, bei2002,bbg2004, brown2006, glt2011,beiersdorfer2017, Shah19}. This suggests measuring other key line formation processes such as dielectronic recombination with better constraints in order to find a plausible explanation for these persistent discrepancies. 

Dielectronic recombination (DR) is the strongest electron-ion recombination process for \ion{Fe}{17} in most photoionized and collisionally ionized astrophysical plasmas \citep{burgess64}, producing satellite lines to the $3d\mbox{ -- }2p$ transition lines through resonant electron capture and subsequent radiative decay. Understanding whether its contributions to \ion{Fe}{17} line formation are causing the model-data discrepancy is essential for improving plasma diagnostics. Validations of these contributions also benchmarking state-of-the-art collisional-radiative models and atomic databases such as SPEX \citep{kaastra1996}, AtomDB \citep{fsb2012}, and CHIANTI \citep{delzanna2015}, which themselves will be needed to interpret observations from the Athena X-IFU \citep{barret2016}, LEM \citep{lem2023}, and Arcus \citep{arcus} high-resolution X-ray imaging spectrometers.


In this work, we use the Flexible Atomic Code (FAC) to  calculate line emission cross sections for the \ion{Fe}{17} DR, DE, and resonant excitation (RE) channels with configurations including principal and orbital angular momentum quantum numbers up to $n\leq7$, $n'\leq100$, and $l,l'\leq8$ respectively, and we benchmark these predictions using experimental cross sections in \ion{Fe}{17} ions that were mono-energetically excited in an electron beam ion trap (EBIT) \citep{Levine88}. In particular, we focus on the cross sections for the higher X-ray energy $n\geq4$ satellites of \ion{Fe}{17} observable in the experimental data. \S~\ref{sec:ebit} describes the EBIT experiment and previous analyses of our measurements. \S~\ref{sec:calc} describes our atomic model calculations. \S~\ref{sec:results} describes the data calibration and shows the theory-experiment comparison and analysis. In \S~\ref{sec:disc} we further discuss our results and future directions.

\section{Experiment}
\label{sec:ebit}

We used FLASH-EBIT \citep{Epp2010} at the Max Planck Institute for Nuclear Physics in Heidelberg, Germany (MPIK) to produce a high-purity ion population mainly consisting of \ion{Fe}{17} ions \citep{Shah19}. A molecular beam of iron pentacarbonyl was injected into the trap through a differentially-pumped injection system, ionized to high charge states by successive electron impact using a mono-energetic and unidirectional electron beam, and compressed by a 6-T magnetic field produced by superconducting Helmholtz coils. The resulting ions were radially trapped by the negative space charge of the compressed electron beam and electrostatically confined in the axial direction by potentials applied to the surrounding cylindrical drift tubes.

For this experiment, the electron-beam energy was swept over the range containing the \ion{Fe}{17} dielectronic capture resonances. The ion population was optimized by applying a charge-breeding time of 0.5 seconds at 1.15 keV, followed by a 40 ms-long ramp-down to 0.3 keV and a symmetric ramp-up. This maximizes the \ion{Fe}{17} purity by efficiently suppressing lower charges states. The electron-beam current was synchronously varied following the relation $n_{e} \propto I_{e}/ \sqrt{E_{e}}$ \citep{2000savin} in order to maintain a constant electron density in the trap. The radiative decay of the excited states generated X-ray photons, which were then collected at $90^{\circ}$ to the electron-beam axis using a silicon-drift detector (SDD) with a photon-energy resolution of $\sim$120 eV FWHM at 1 keV. The unidirectional electron beam causes anisotropic, polarized X-ray emission from the trapped ions \citep{beiersdorfer1996, shah2015, Shah18}.

In previous works, we measured line emission cross sections for the $3s\mbox{ -- }2p$ and $3d\mbox{ -- }2p$ channels of \ion{Fe}{17} ions formed through DR, RE, DE, and radiative cascades following RR, as well as intensities and cross sections of Fe DR L-shell satellites for the $LMn$ ($2p^5 3l nl' \rightarrow 2p^6 3l$) series for \ion{Fe}{17} \citep{Shah19}. These measurements improved on previous experiments \citep{B98, beiersdorfer2017} by reducing the collision-energy spread to only 5\,eV full-width-at-half-maximum (FWHM) at 800 eV. We also obtained experimental resonant strengths and rate coefficients for the DR $LMM$ (3$l$3$l'$) satellites of \ion{Fe}{17} \citep{Grilo2021}. The calculated rate coefficients were compared with those available in the OPEN-ADAS and AtomDB \citep{fsb2012} databases, both of which are frequently used in spectral analyses, ultimately unveiling disparities of $9-12\%$ and $\sim$5$\%$ respectively.

\begin{figure*}[h]
\hspace*{-0.75cm}
\includegraphics[height=0.5 \textwidth, width = 1.2 \textwidth]{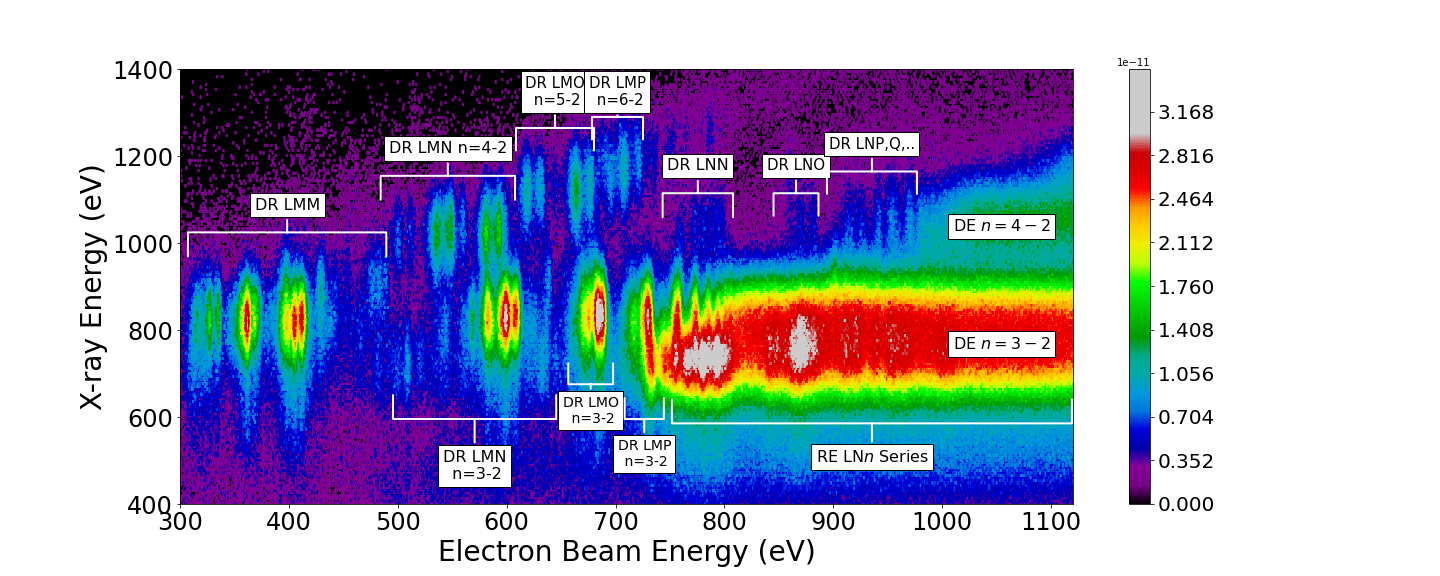}

\hspace*{-0.75cm}
\includegraphics[height=0.5 \textwidth, width=1.2 \textwidth]{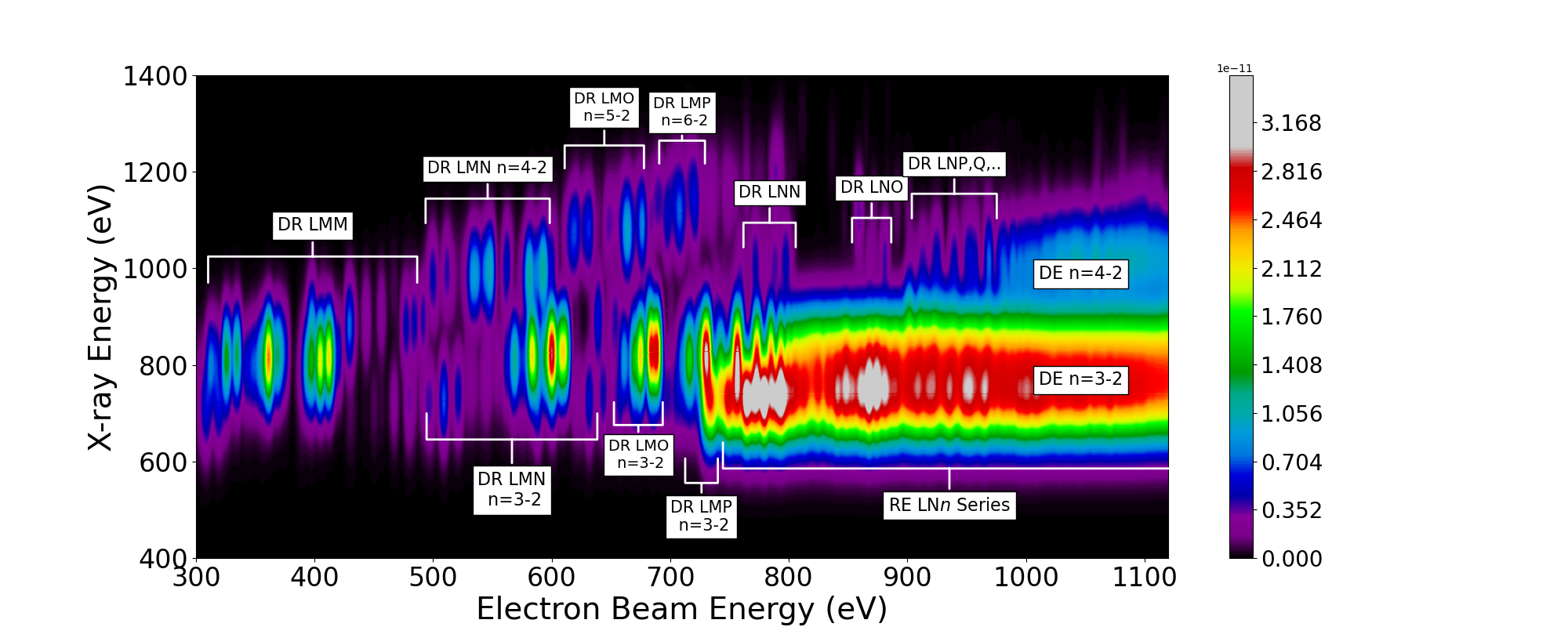}

\caption{X-ray photon flux from the FLASH-EBIT experiment (top) and FAC calculations (bottom) as a function of photon energy and electron-beam energy. The labelled features represent resonances formed through dielectronic recombination (DR), resonant excitation (RE), and direct electron-impact excitation (DE) at \ion{Fe}{17} transition lines. The $LMn$ and $LNn$ series represent the 3$l$n$l'$ and 4$l$n$l'$ L-shell satellites of \ion{Fe}{17} respectively. The color scale represents the line-flux intensity. For direct comparison with the experimental spectrum, FAC-calculated cross sections were folded along the electron-beam energy and photon-energy axes to match the resolution of the nearly mono-energetic electron beam of FLASH-EBIT ($\sim$5 eV FWHM) and the photon-energy resolution of the used silicon-drift detector ($\sim$120 eV FWHM).
\label{fig:contours}}
\end{figure*}

\section{Electronic-structure calculations}
\label{sec:calc}

We extend the work of \cite{Shah19} by calculating cross sections for all DR satellite lines of \ion{Fe}{17} observable in the FLASH-EBIT experiment. We used the Flexible Atomic Code (FAC) \citep{Gu2008} to obtain the electronic structure for the initial, intermediate, and final states of \ion{Fe}{17} ions, as well as their transition and autoionization rates. In order to match the polarized experimental emission, we fed these atomic data into the line-polarization module of FAC (FAC-$\it{pol}$), which computes line polarizations resulting from the uni-directional electron beam, to then calculate the differential (observed at 90$^{\circ}$) and total line emission cross sections for each region-of-interest (ROI).

We performed calculations for the dielectronic capture channels of DR, RE, and DE by including $2s^2 2p^5 nl n'l'$ configurations with principal quantum numbers and orbital angular momentum quantum numbers up to $n \leq 7$, $n' \leq 100$, and $l,l' \leq 8$ respectively, and allowing full-order configuration mixing. In doing so, we accounted for all DR $LMn$ and $LNn$ ($2p^5 4l nl' \rightarrow 2p^6 4l$) satellites resolvable in our experiment. We calculated the DR resonant strengths in the isolated resonance approximation as in previous studies \citep{Shah19, Grilo2021}, meaning we assumed no quantum interference between DR resonances or with non-resonant recombination channels \citep{pindzola2006, Zatsarinny2005}. In this approximation, the DR strength is 

\begin{equation}
S^{DR}_{idf} = \int_{0}^{\infty} \sigma^{DR}_{idf}(E_{e})dE_{e} = \frac{\pi^2 \hbar^3}{m_{e}E_{id}} \frac{g_{d}}{2g_{i}} \frac{A^{a}_{di} A^{r}_{df}}{\Sigma_{i'} A^{a}_{di'} + \Sigma_{f'} A^{r}_{df'}}
\label{eqn:dr}
\end{equation}

\noindent where $\sigma^{DR}_{idf}(E_{e})$ is the DR cross section as a function of the free-electron kinetic energy $E_{e}$, $m_{e}$ is the electron mass (in units MeV/$c^2$), $\hbar$ is the reduced Planck constant, and $E_{id}$ is the resonant energy of the electron-ion recombination between the initial state $i$ and intermediate doubly excited state $d$ with statistical weights $g_{i}$ and $g_{d}$. $A^{a}_{di}$ and $A^{r}_{df}$ represent the autoionization rate between states $d$ to $i$ and radiative transition rate between state $d$ and the final state $f$ respectively, which were both calculated with FAC.

\section{Data Analysis and Results}
\label{sec:results}

\subsection{Data Calibration}
\label{subsec:gain}

We calibrated the experimental data first by correcting for the filter transmission resulting from the 1 $\mu$m carbon foil in front of the SDD, which shields it from UV light but also blocks a part of the X-ray radiation from the trap. We calculated the transmission using optical constants from \cite{hanke1993}. \cite{Shah19} verified the filter transmission through Ly$\alpha$ and radiative recombination emission measurements of well-known ions \ion{O}{8} and \ion{Ne}{10}, finding agreement within $3\%$. We include this uncertainty in our error budget. 

As described in \S~\ref{sec:ebit}, the beam current $I_{e}$ was adjusted while the electron-beam energy $E_{e}$ was changed in order to maintain a constant electron density $n_{e}$. Since the X-ray intensity is proportional to the electron beam current density $j_{e}$, which in turn depends on the product of beam current and square root its energy \citep{1995wong}. Therefore, we corrected the observed X-ray count rate by dividing by a factor of $\sqrt{E_{e}}$.

We also calculated a correction for the nominal SDD energy scale. Using calculated centroid-photon energies for both the FAC-calculated and FLASH-measured resonances, we fit a linear model to calculate the gain correction. We used the centroid-photon energies for the 3$s$, 3$d$, and 4$d$ manifolds for the linear fit in addition to the origin. We calibrated the electron-beam energy using the $LMM$ and $LMN$ $n=3-2$ DR resonant energies \citep{bei2014}, both of which are theoretically well known.

\begin{figure*}[h]
\epsscale{1.0}
\vspace*{-1cm}\plotone{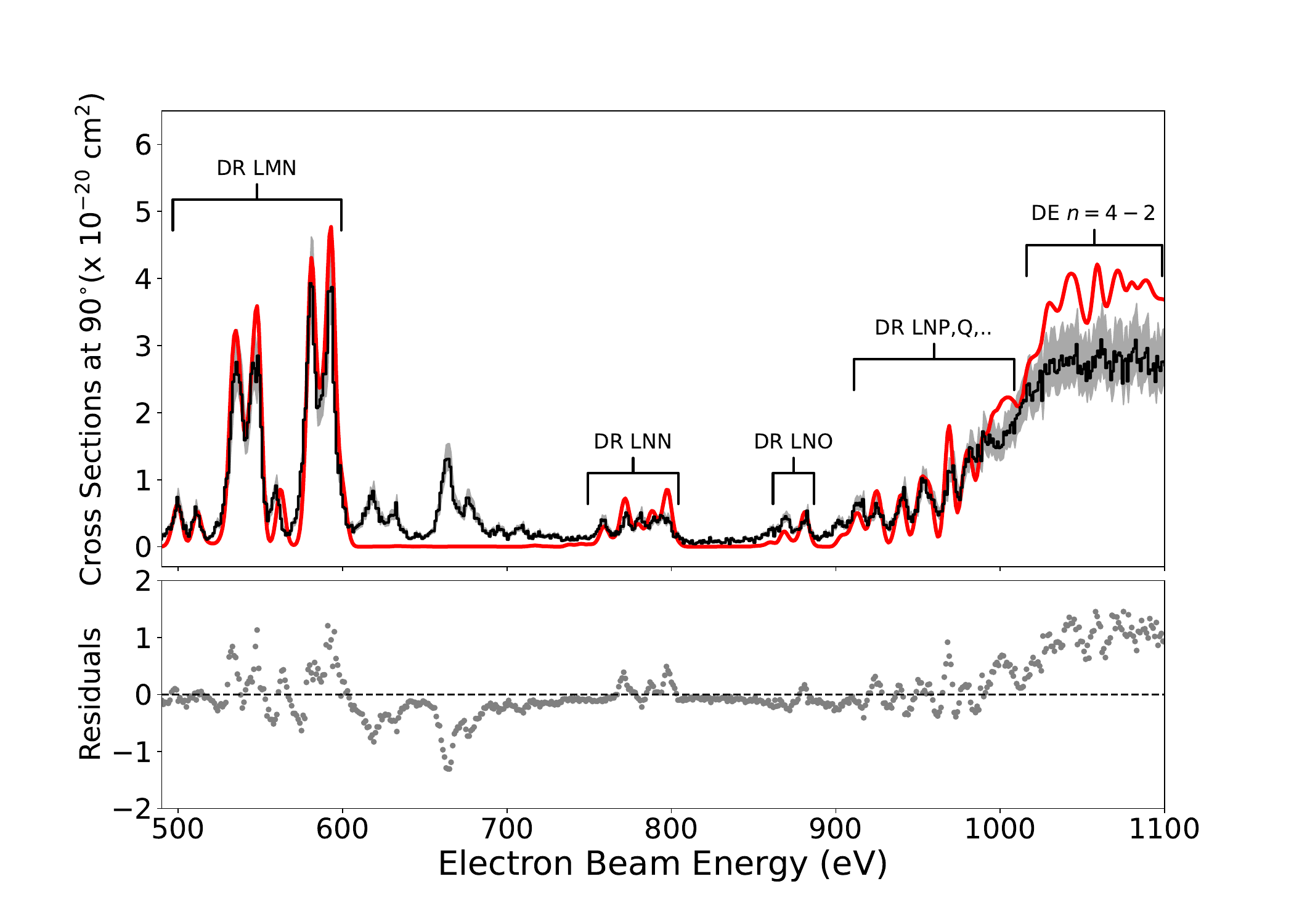}\vspace*{-1cm}
\plotone{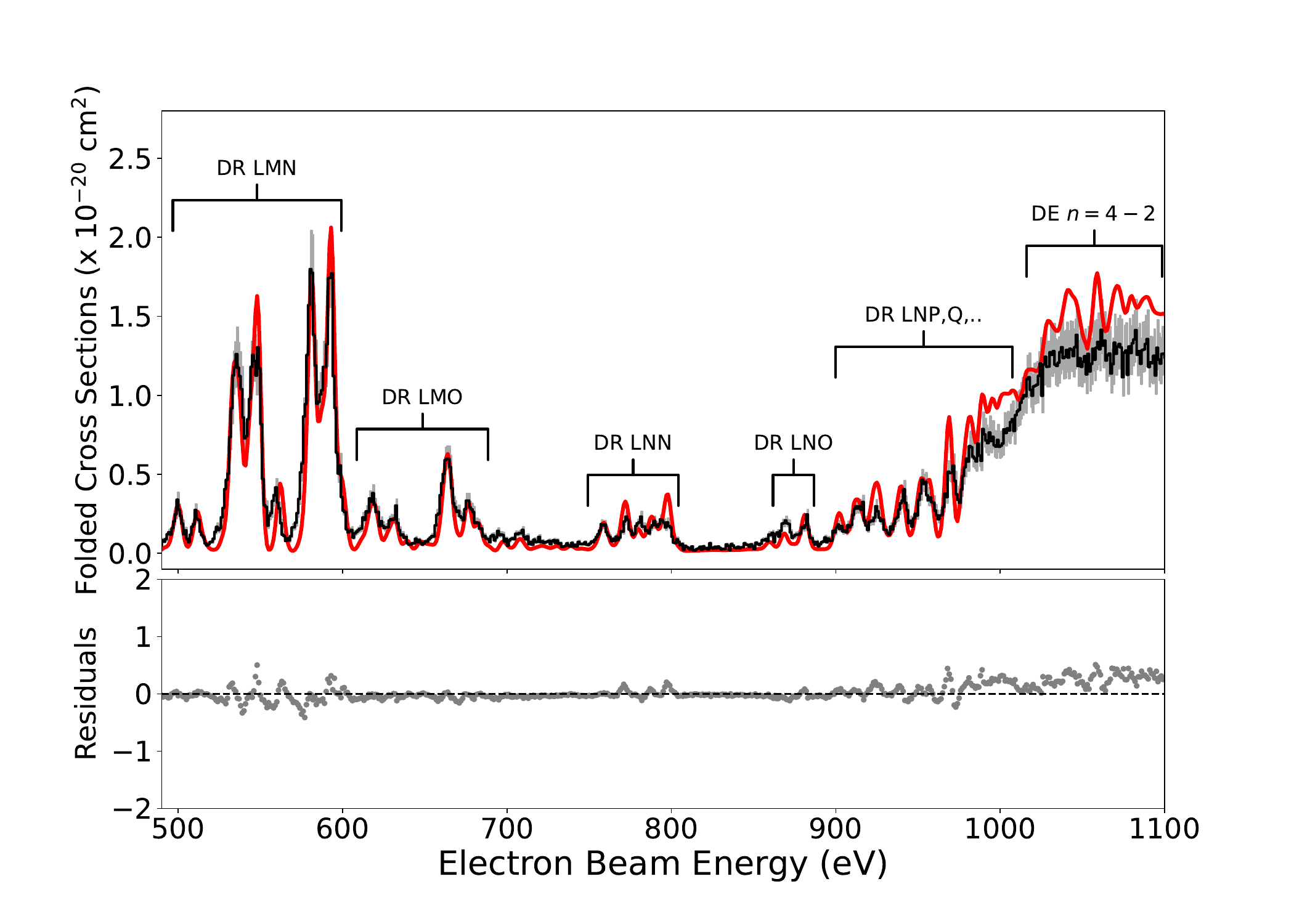}
\caption{Top: Experimental cross sections observed at 90$^{\circ}$ (black curve) versus FAC-calculated absolute cross sections (red curve) as a function of electron-beam energy for the \ion{Fe}{17} $LMN$ $n=4-2$ channels within the 980 - 1030 eV photon-energy range, revealing an overestimation on the scale of $20-25\%$ for both the DR and DE channels. Bottom: Same as above, except for forward-folded theoretical FAC cross sections (red), which agree much better overall with the DR and DE channels. The features missing in the absolute cross section plot (mostly due to LMO transitions) are now visible after matching the photon-energy resolution of the FLASH-EBIT silicon-drift detector. The shaded gray bands represent the range of uncertainty (systematic and statistical) for the experimental cross sections.
\label{fig:n=4}}
\end{figure*}

\begin{figure}[ht]
\epsscale{1.3}
\hspace*{-0.4cm}\plotone{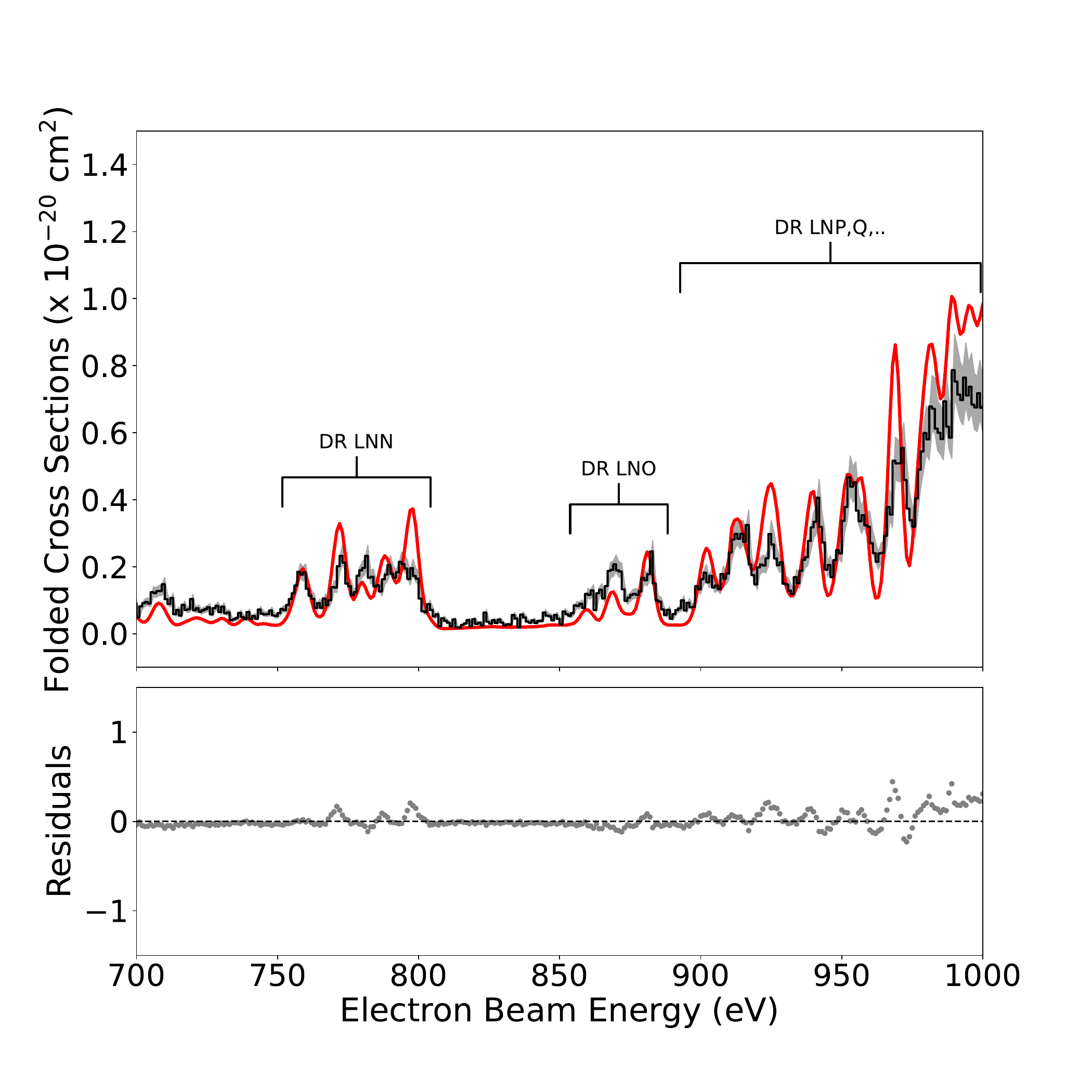}
\vspace*{-0.5cm}
\caption{Forward-folded theoretical (red) versus experimental (black) DR cross sections observed at 90$^{\circ}$ for the \ion{Fe}{17} $LNn$ series within the 980 - 1030 eV photon-energy range. 
\label{fig:dr_LNN_smear}}
\end{figure}

\begin{figure}[ht]
\epsscale{1.3}
\hspace*{-4mm}\plotone{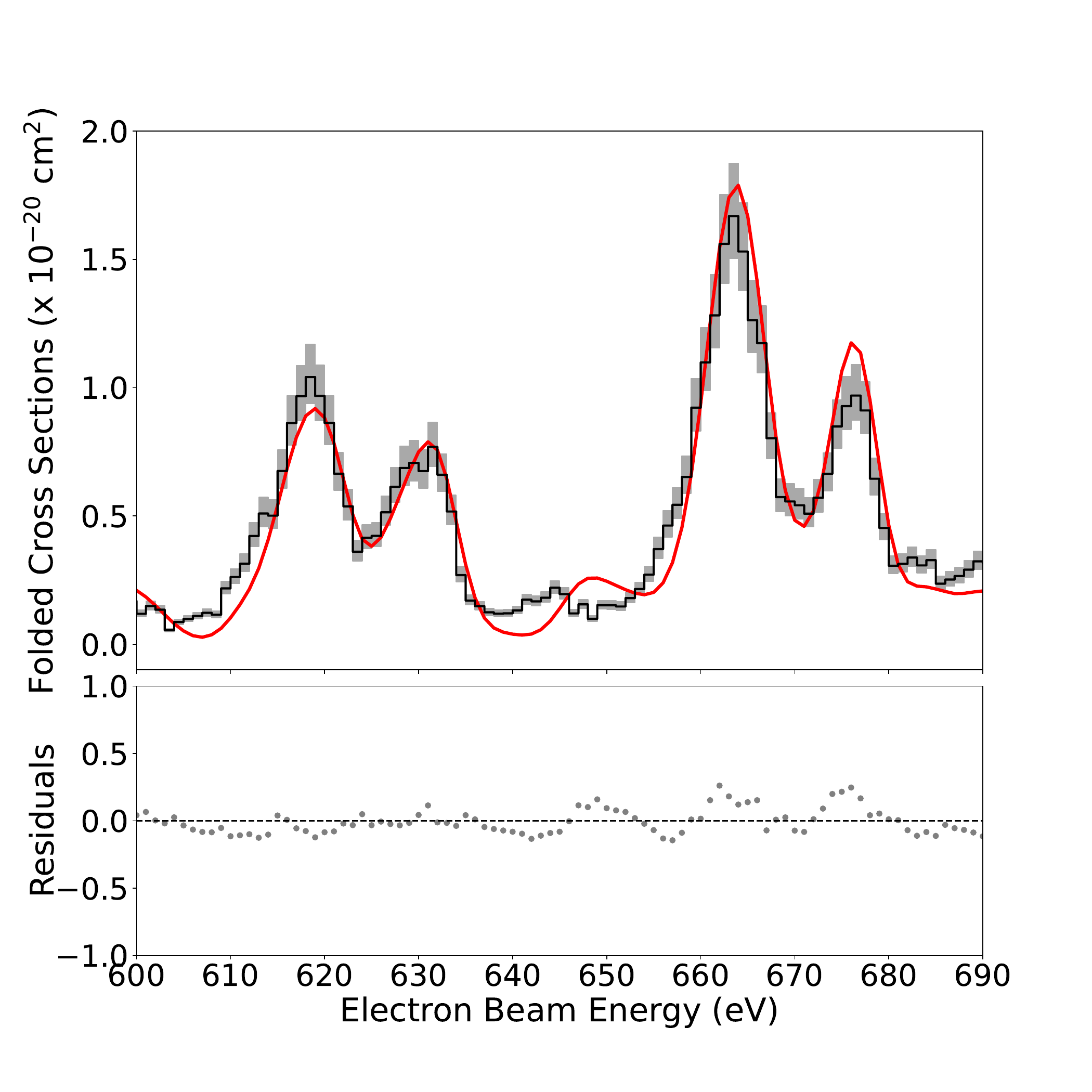}
\vspace*{-0.5cm}
\caption{Forward-folded theoretical (red) versus experimental (black) DR cross sections observed at 90$^{\circ}$ for the \ion{Fe}{17} $LMO$ $n=5-2$ resonances within the 1050 - 1120 eV photon-energy range. 
\label{fig:dr_lmo_smear}}
\end{figure}

\begin{figure}[ht]
\epsscale{1.3}
\plotone{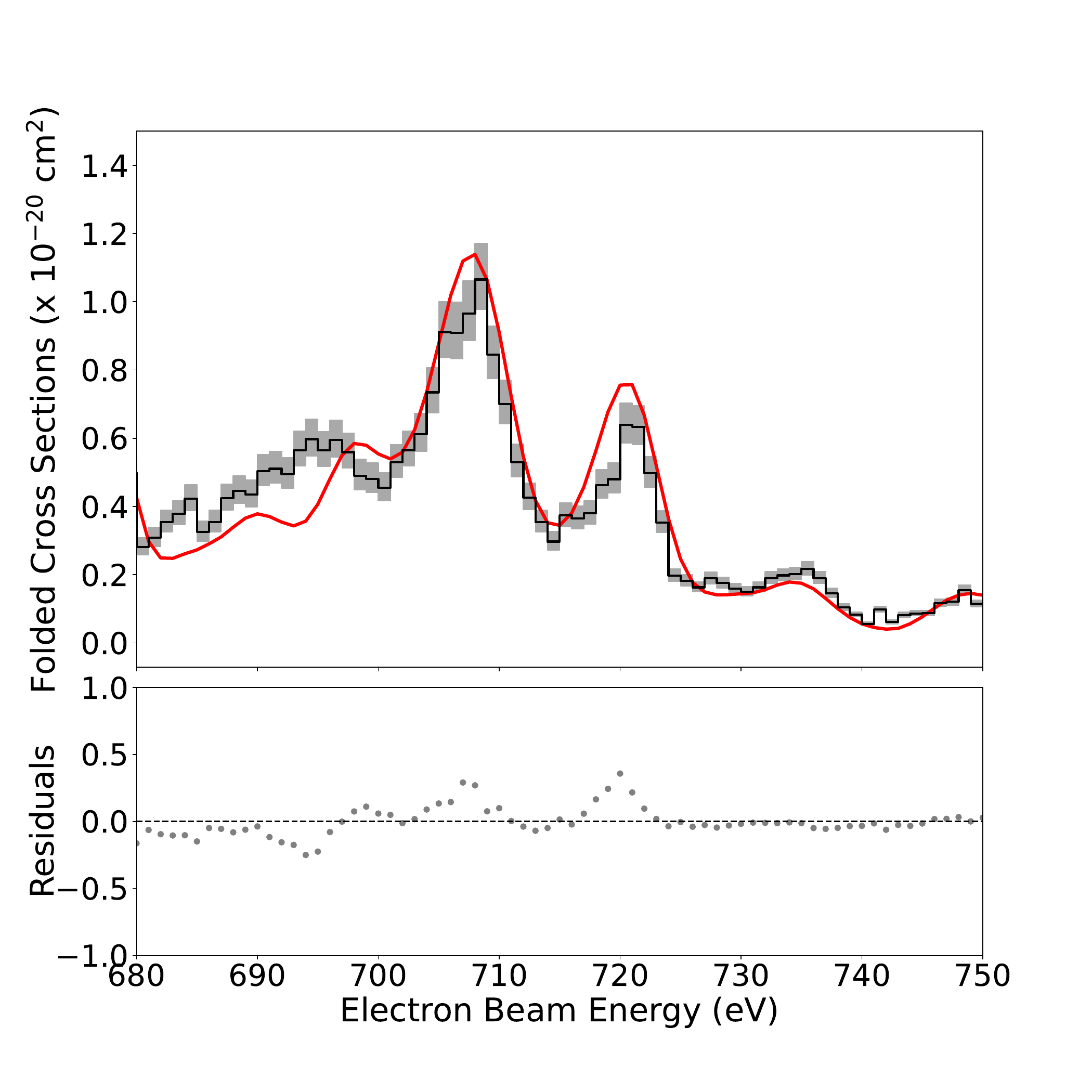}
\vspace*{-0.5cm}
\caption{Forward-folded theoretical (red) versus experimental (black) DR cross sections observed at 90$^{\circ}$ for the \ion{Fe}{17} $LMP$ $n=6-2$ resonances within the 1090 - 1170 eV photon-energy range. 
\label{fig:dr_lmp_smear}}
\end{figure}

\subsection{Cross Section Calibration}
\label{subsec:cscomp}

For $n=3$, we selected a single DR resonance at 412 eV electron beam energy to normalize the experimental counts to our theoretical cross sections, as in \cite{Shah19}. We normalized the experimental counts for the $n\geq4$ complexes using similar criteria. We selected a single DR resonance at 511 eV electron beam energy for $n=4$, as its strength can be traced to a single strong resonance in the $LMN$ $n=4-2$ channel $[((2p_{1/2} 2p^{4}_{3/2})_{1/2} 3s_{1/2})_{1} 4d_{3/2}]_{3/2}$. Similarly, we used the $[((2p_{1/2} 2p^{4}_{3/2})_{1/2} 3d_{5/2})_{2} 5d_{3/2}]_{7/2}$ channel at 677 eV electron energy to normalize the $LMO$ $n=5-2$ cross sections, and the $[((2p_{1/2} 2p^{4}_{3/2})_{1/2} 3d_{5/2})_{3} 6d_{3/2}]_{7/2}$ channel at 721 eV for the $LMP$ $n=6-2$ cross sections.

However, due to the finite energy resolution of the SDD, when selecting a ROI on the experimental data, the resulting spectral histogram loses some flux from the resonances of interest while gaining some flux from features meant to be excluded. This effect becomes more prominent at $n\geq4$, as the DR resonances closely overlap along the photon-energy axis. 

\subsection{Forward-folded Cross Sections}
\label{subsec:ffcomp}

To overcome the issue of finite detector resolution impacting the spectrum resulting from an ROI selection on the photon-energy axis, we employed a "forward-folding" approach. In our previous work \citep{Shah19, Grilo2021}, we broadened the theoretical cross sections only on the electron beam energy axis, accounting for its Gaussian distribution, which dominates the broadening. In the present approach, we also applied Gaussian broadening to the cross sections along the photon-energy axis to match the finite detector resolution, thereby folding the theoretical cross sections with the experimental response function. Because other broadening terms, such as natural linewidth and Doppler broadening, are negligible compared with the electron beam energy spread and photon-detector resolution, we only included these two dominant terms, so that 
\begin{equation}
    \sigma_i (E_e, E_\gamma) = \sigma_{i,total}\, G(E_e, E_{e,0}, w_e)\, G(E_\gamma, E_{\gamma,0}, w_\gamma)
\end{equation}
where $\sigma_i$ is the cross section of process $i$; $E_e$ and $E_\gamma$ are the electron beam and photon energies, respectively; $\sigma_{i,total}$ is the total cross section for process $i$; and $G(E, E_0, w_i)$ are Gaussian distributions centered on $E_0$, with widths $w_{e}$, $w_{\gamma}$ representing the FWHM of electron-energy spread and photon-detector resolution respectively, and evaluated at $E$. 

By forward-folding the theoretical cross sections through the detector response before applying the ROI cut, we are able to reproduce blended features that are present in the experimental spectrum, thus getting better fidelity in the direct comparison of the higher $n$ cross sections. This approach is not as useful when attempting to separate closely spaced line manifolds such as $3d\mbox{ -- }2p$ and $3d\mbox{ -- }2p$, as they are very sensitive to the exact choice of ROI.

The electron beam energy resolution was set to match the energy spread of the quasi-mono-energetic electron beam ($w_e$ = 5 eV) of FLASH-EBIT. The X-ray photon-energy resolution $w_\gamma$ was set to $\sim$120 eV to match that of the SDD. Figure~\ref{fig:contours} shows two-dimensional contour plots comparing both the measured X-ray flux from the FLASH-EBIT experiment (top panel) as a function of electron beam energy and photon energy and the now-folded FAC-calculated X-ray intensity (bottom panel).

With the forward-folding approach, we now determine a normalization factor by computing the amplitudes of the 412-eV peak for both the experimental and folded theoretical projections (with the same ROI selections) and taking the quotient. In this way we effectively reduced the systematic error stemming from the ROI selection bias from $9\%$ down to $3\%$. The uncertainty (statistical and systematic) sources from counting statistics, the carbon-foil transmission correction, and the normalization factor yielded total estimates of $12\%$, $11\%$, and $9\%$ for the $n=4$, $n=5$, and $n=6$ forward-folded experimental cross sections respectively, which are all $\sim$2-3$\%$ improvements from the absolute cross section uncertainties. In the case of the $n\geq4$ complexes, we are now primarily limited by counting statistics, since weaker lines show larger statistical uncertainties.

For the $n=4-2$ $LMN$ cross-section comparison, we selected an ROI along the photon-energy axis ranging from 980 - 1030 eV in order to include every relevant DR and DE contribution in the complex. We then plotted a one-dimensional histogram of this ROI to represent the differential cross sections  as a function of electron-beam energy. Figure~\ref{fig:n=4} shows a comparison between both the absolute theoretical DR cross sections (top panel) and forward-folded cross sections (bottom panel) versus the experimental data. The forward-folded cross sections agree much better with the experimental results and exhibit previously hidden features that blend in due to the energy resolution of the SDD. Figure~\ref{fig:dr_LNN_smear} shows a zoom in on the forward-folded versus experimental cross sections for the DR $LNn$ series, which is also included in the same photon-energy ROI.

For the $n=5-2$ $LMO$ and $n=6-2$ $LMP$ cross-section comparisons, we selected horizontal ROIs of 1050-1120 eV and 1090-1170 eV X-ray energy range respectively. Wider ROIs were necessary for these channels in order to include all DR resonances. Figures~\ref{fig:dr_lmo_smear} and \ref{fig:dr_lmp_smear} show the forward-folded versus experimental cross-section comparisons for these complexes.

\subsection{Rate Coefficients}
\label{sec:rates}

We tabulate the integrated resonant strengths in Table \ref{tab:res_strengths} for each defined electron beam energy region. Because we cannot get individual DR resonant strengths due to blending, we integrated over beam-energy ranges. We also inferred rate coefficients from both the experimental results and the FAC-calculated absolute DR total cross sections, as they are convenient parameters for spectral modelling as well as collisional-radiative models of single-temperature and multi-temperature astrophysical plasmas. 

As in \cite{Grilo2021}, we converted the experimental cross sections observed at $90^{\circ}$ (with respect to the electron beam) to total cross sections using the formula $S^{total} = 4\pi I^{90^{\circ}} / W(90^{\circ})$, where $I^{90^{\circ}}$ represents the observed DR intensity and  $W(90^{\circ}) = 3/(3 - P)$ is a polarization correction factor in which P is the polarization for a specific radiative transition \citep{beiersdorfer1996}. We estimated total uncertainties of $10\%$, $15\%$, $13\%$, and $11\%$ for the $n=3$, $n=4$, $n=5$, and $n=6$ experimental resonant strengths respectively from the counting statistics, carbon-foil transmission correction, and normalization factor calculations of the absolute (i.e. not forward-folded) cross sections. These uncertainties appear in the rate coefficient conversion. 

\begin{table}[ht]
  \caption{Experimental and FAC-calculated integrated cross sections ($10^{-20} \text{cm}^{2}$ eV) with deviations (relative data-model disagreement $\pm$ experimental uncertainty).
\label{tab:res_strengths}}
    \hspace*{-1.0cm}\begin{tabular}{c|cccc}
    \hline \hline
      Channel & Energy & $S_{FLASH}$ & \multicolumn{2}{c}{$S_{FAC}$}\\ 
       & (eV) &  &  FAC & Deviation
      \\ \hline
      $LMM$ (3s$\rightarrow$2p) & 300-340 & 85 $\pm$ 10 & 80.62 & ($6\%$ $\pm$ $13\%$)\\
      $LMM$ (3p$\rightarrow$2p) & 340-380 & 215 $\pm$ 30 & 199.37 & ($8\%$ $\pm$ $16\%$))\\
      $LMM$ (3d$\rightarrow$2p) & 380-420 & 300 $\pm$ 40 & 263.79 &($14\%$ $\pm$ $15\%$)\\
      $LMN$ (3{\it l}$\rightarrow$2p) & 560-620 & 440 $\pm$ 60 & 508.38 & ($-15\%$ $\pm$ $16\%$)\\
      $LMN$ (4{\it l}$\rightarrow$2p) & 490-610 & 115 $\pm$ 15 & 137.91 & ($-20\%$ $\pm$ $15\%$)\\
      $LMO$ (3{\it l}$\rightarrow$2p) & 650-700 & 370 $\pm$ 60 & 423.56 & ($-14\%$ $\pm$ $20\%$)\\
      $LMO$ (5{\it l}$\rightarrow$2p) & 740-810 & 70 $\pm$ 10 & 71.50 & ($-3\%$ $\pm$ $17\%$)\\
      $LMP$ (3{\it l}$\rightarrow$2p) & 700-750 & 320 $\pm$ 50 & 339.93 & ($-7\%$ $\pm$ $19\%$)\\
      $LMP$ (6{\it l}$\rightarrow$2p) & 850-890 & 30 $\pm$ 5 & 35.56 & ($-18\%$ $\pm$ $20\%$)\\
      $LNN$ (total) & 740-810 & 16 $\pm$ 2 & 19.10 & ($-19\%$ $\pm$ $15\%$)\\
      $LNO$ (total) & 850-890 & 9 $\pm$ 1 & 4.86 & ($85\%$ $\pm$ $13\%$)\\
      \hline \hline
    \end{tabular}
\end{table}

The DR rate coefficients were obtained by integrating the corresponding DR resonant strengths over a Maxwellian velocity distribution of the electrons \citep{Gu2003} as shown in Equation \ref{eqn:dr_rate} below

\begin{equation}
\alpha^{DR}_{if} = \frac{m_{e}}{\sqrt{\pi}\hbar^3} \left(\frac{4 R_{\infty}}{k_{B} T_{e}}\right)^{3/2} a_{0}^{3} \sum_{d} E_{id} \,  S^{DR}_{idf} \, {\exp \left( -\frac{E_{id}}{k_{B} T_{e}}\right)}\, ,
\label{eqn:dr_rate}
\end{equation}
where $R_{\infty}$ is the Rydberg constant in eV, $a_0$ is the Bohr radius, $k_{B}$ is the Boltzmann constant, and $T_{e}$ is the electron temperature. We provide the inferred experimental and theoretical rate coefficients in Table \ref{tab:dr_rates} for each defined electron-beam energy region at different plasma-electron temperatures. For comparison, we include DR rates reported in the OPEN-ADAS online atomic database. The electron temperatures of 110.3\,eV and 220.3\,eV were used for direct comparison with the tabulated DR rates retrieved from OPEN-ADAS (files: {\tt nrb00\#ne\_fe16ls24.dat} (nrbLS), {\tt nrb00\#ne\_fe16ic24.dat} (nrbIC)). These calculations were provided by author N. Badnell in both LS and intermediate couplings (IC).

\section{Discussion}
\label{sec:disc}

There are several noticeable discrepancies between the $n=4$ absolute theoretical cross sections and experimental data in the top panel of Figure~\ref{fig:n=4}. Most prominently, there is a considerable overestimation of the $n=4-2$ $LMN$ $4p$ (0.535, 0.547 keV) and $4d$ (0.581, 0.592 keV) DR channels by $20\%$, as well as the DE cross sections in the beam-energy range of 1.0-1.1\,keV electron energy by $25\%$. The uncertainty for the $n\geq4$ channels is also large, as shown by the shaded gray band.

However, in the bottom panel we see improved overall agreement between the experimental and forward-folded theoretical cross sections for the $n=4-2$ comparisons over a wide range of electron energies, particularly for the DR channels at 0.535 and 0.592\,keV. The discrepancies for the other resonances as well as the DE cross sections are now reduced to $\leq15\%$. Additionally, the lower features observed between 0.6-0.75\,keV, which are due to the high-energy tail of the $LMO$ DR resonances, are now visible and match the experimental data. We also reduce the total uncertainty by $3\%$ on account of the lower systematic error from the ROI selection.

Despite the improved agreement, there are still a few noticeable discrepancies in the forward-folded comparisons. For the $LNn$ cross sections in Figure~\ref{fig:dr_LNN_smear}, the observed peak at 0.797 keV is smaller than predicted. The predicted $LNO$ peaks (blends of $5p$, $5d$, and $5f$) at 0.859 and 0.868 keV are both slightly smaller than the measured ones, which contributes to the disparity with the calculated resonant strengths and rate coefficients for this region.

Similar levels of agreement are observed for the $LMO$ forward-folded cross sections. The $LMO$ DR channels observed at 0.619, 0.631, and 0.664 keV in Figure~\ref{fig:dr_lmo_smear} all agree within $\leq10\%$, though the DR channel at 0.677 keV exhibits a $20\%$ discrepancy. The $LMP$ DR channels observed at 0.698 and 0.708 keV seen in Figure~\ref{fig:dr_lmp_smear} also agree within $10\%$, though we notice $20\%$ overestimations for the channels at 0.690 and 0.721 keV. It appears that we may be missing contributions in our calculations for the resonances observed between 0.68-0.7\,keV. A few DR features also appear shifted by 1-2\,eV, particularly the channels observed at 0.648\,keV in $LMO$ and 0.721\,keV in $LMP$.

We reasonably agree within $2\sigma$ when comparing most of the experimentally estimated rate coefficients to both our predictions and data compiled in OPEN-ADAS. Disparities become more noticeable for the higher $n$ complexes, particularly the $LNn$ series. However, the total integrated cross sections are smaller for these resonances, and therefore less consequential. 

Regarding other atomic databases, neither do we compare our results to AtomDB, as the rates for the $n\geq4$ channels are not yet available, nor with SPEX, as the \ion{Fe}{17} rates available are from FAC calculations done by \cite{Shah19,gu2020} which made nearly identical calculations to ours for up to $n'\leq60$, what makes such a comparison pointless.

Determining accurate rate coefficients for these DR satellite lines is therefore crucial for reliable diagnositcs of hot astrophysical plasmas. This has been shown e.~g., in Refs. \citep{gabriel72, beiersdorfer2018}, where plasma-electron temperatures in range 0.2 - 0.6 keV were obtained from the intensity ratio of the satellite lines over the 3$C$ resonance transition.

\section{Conclusions}

In the present work, we compared our dedicated FAC cross-section calculations for the dielectronic recombination satellites of \ion{Fe}{17} to those extracted from our experiment using FLASH-EBIT. We thereby extended the experimental benchmark to higher electron energies to excite $n\geq4$ DR resonances of \ion{Fe}{17}. We improved on previous work by applying a "forward-folding" approach in which we broadened our theoretical cross sections to match the photon-energy resolution and the experimental width of the electron-beam energy. Moreover, by combining our calculations and experimental data, we inferred DR rate coefficients. This allows us to benchmark those compiled in the OPEN-ADAS database, which were found to agree within $2\sigma$ with our improved results.

Performing the same experiment with a high-resolution wide-band X-ray microcalorimeter instead of an SDD would much reduce systematic uncertainties on our DR cross sections. That instrument would enable a far more clear selection of regions of interest in the data, in most cases encompassing individual resonances. Improving DR cross sections for those is a critical task in the perspective of the wealth of observational data expected from X-ray observations with XRISM, ATHENA, LEM, and Arcus, as we will not be able to extract the full diagnostic information that their high-resolution would afford without experimentally benchmarked atomic data for collisional excitation cross sections and DR rates.

\acknowledgements

We acknowledge support from NASA's Astrophysics Program. G.G. acknowledges support under NASA award No. 80GSFC21M0002. P.~A.~acknowledges the support from Funda\c{c}\~{a}o para a Ci\^{e}ncia e a Tec\-no\-lo\-gia (FCT), Portugal, under Grant No. UID/FIS/04559/2020(LIBPhys) and High Performance Computing Chair - a R\&D infrastructure based at the University of Évora (PI: M. Avillez). C.S. acknowledges support from NASA under award number 80GSFC21M0002 and by the Max-Planck-Gesellschaft (MPG).

\clearpage

\begin{longtable*}{c|cccc|cccc}
\caption{FAC-calculated rate coefficients ($10^{-13}$ cm$^{3}$ s$^{-1}$) for different electron temperatures (eV) compared to those reported from OPEN-ADAS. \label{tab:dr_rates}}
    \\ \hline\hline
      Channel & $T_e$ & FLASH & \multicolumn{2}{c}{FAC} & \multicolumn{4}{c}{OPEN-ADAS} \\ 
       & & & FAC & Deviation & nrbLS & Deviation & nrbIC & Deviation
      \\ \hline
      $LMM$  & 110.3 & 44 $\pm$ 6 & 39.1 & ($12\%$ $\pm$ $16\%$) & 43.3 & ($2\%$ $\pm$ $16\%$) & 41.2 & ($7\%$ $\pm$ $16\%$)\\
      (total) & 220.3 & 84 $\pm$ 10 & 75.3 & ($11\%$ $\pm$ $14\%$) & 95.2 &($-12\%$ $\pm$ $14\%$) & 96.2 & ($-13\%$ $\pm$ $14\%$)\\
      & 300 & 83 $\pm$ 10 & 74.7 & ($11\%$ $\pm$ $14\%$) & - & - & - & -\\
      & 2000 & 14 $\pm$ 2 & 12.7 & ($10\%$ $\pm$ $17\%$) & - & - & - & -\\
      \hline
      $LMN$  & 110.3 & 6.9 $\pm$ 1 & 7.87 & ($-14\%$ $\pm$ $17\%$) & 7.27 & ($-6\%$ $\pm$ $17\%$) & 7.27 & ($-6\%$ $\pm$ $17\%$)\\
      (3{\it l}$\rightarrow$2p) & 220.3 & 36 $\pm$ 5 & 41.3 & ($-15\%$ $\pm$ $16\%$) & 38.3 & ($-7\%$ $\pm$ $16\%$) & 39.5 & ($-10\%$ $\pm$ $16\%$)\\
      & 300 & 46 $\pm$ 6 & 53.4 & ($-14\%$ $\pm$ $15\%$) & - & - & - & -\\
      & 2000 & 14.5 $\pm$ 2 & 16.8 & ($-16\%$ $\pm$ $16\%$) & - & - & - & -\\
      \hline
      $LMN$  & 110.3 & 2.4 $\pm$ 0.3 & 2.73 & ($-14\%$ $\pm$ $15\%$) & - & - & - & -\\
      (4{\it l}$\rightarrow$2p) & 220.3 & 10.5 $\pm$ 1.2 & 12.3 & ($-17\%$ $\pm$ $13\%$) & - & - & - & -\\
      & 300 & 13 $\pm$ 1.5 & 15.2 & ($-17\%$ $\pm$ $13\%$) & - & - & - & -\\
      & 2000 & 3.7 $\pm$ 0.5 & 4.38 & ($-18\%$ $\pm$ $16\%$) & - & - & - & -\\
      \hline
      $LMO$  & 110.3 & 3.1 $\pm$ 0.5 & 3.46 & ($-12\%$ $\pm$ $20\%$) & 2.94 & ($5.4\%$ $\pm$ $20\%$) & 2.98 & ($4\%$ $\pm$ $20\%$)\\
      (3{\it l}$\rightarrow$2p) & 220.3 & 24 $\pm$ 3 & 26.74 & ($-11\%$ $\pm$ $15\%$) & 22.9 & ($5\%$ $\pm$ $15\%$) & 23.7 & ($2\%$ $\pm$ $15\%$)\\
      & 300 & 34 $\pm$ 4 & 38.27 & ($-13\%$ $\pm$ $14\%$) & - & - & - & -\\
      & 2000 & 13.5 $\pm$ 2 & 15.32 & ($-13\%$ $\pm$ $18\%$) & - & - & - & -\\
      \hline
      $LMO$  & 110.3 & 0.7 $\pm$ 0.15 & 0.75 & ($-7\%$ $\pm$ $28\%$) & - & - & - & -\\
      (5{\it l}$\rightarrow$2p) & 220.3 & 5 $\pm$ 1 & 4.97 & ($1\%$ $\pm$ $25\%$) & - & - & - & -\\
      & 300 & 6.8 $\pm$ 1 & 6.84 & ($-1\%$ $\pm$ $18\%$) & - & - & - & -\\
      & 2000 & 2.5 $\pm$ 0.5 & 2.51 & ($-1\%$ $\pm$ $25\%$) & - & - & - & -\\
      \hline
      $LMP$  & 110.3 & 1.8 $\pm$ 0.2 & 1.95 & ($-9\%$ $\pm$ $13\%$) & 1.52 & ($18\%$ $\pm$ $13\%$) & 1.48 & ($22\%$ $\pm$ $13\%$)\\
      (3{\it l}$\rightarrow$2p) & 220.3 & 17.5 $\pm$ 2 & 18.6 & ($-7\%$ $\pm$ $13\%$) & 14.4 & ($22\%$ $\pm$ $13\%$) & 14.4 & ($22\%$ $\pm$ $13\%$)\\
      & 300 & 26.5 $\pm$ 4 & 28.1 & ($6\%$ $\pm$ $18\%$) & - & - & - & -\\
      & 2000 & 12 $\pm$ 2 & 12.8 & ($7\%$ $\pm$ $20\%$) & - & - & - & -\\
      \hline
      $LMP$  & 110.3 & 0.2 $\pm$ 0.05 & 0.24 & ($-20\%$ $\pm$ $33\%$) & - & - & - & -\\
      (6{\it l}$\rightarrow$2p) & 220.3 & 1.8 $\pm$ 0.2 & 2.08 & ($-16\%$ $\pm$ $13\%$) & - & -  & - & - \\
      & 300 & 2.7 $\pm$ 0.3 & 3.06 & ($-14\%$ $\pm$ $13\%$) & - & - & - & -\\
      & 2000 & 1.1 $\pm$ 0.3 & 1.32 & ($-20\%$ $\pm$ $38\%$) & - & - & - & -\\
      \hline
      $LNN$ & 110.3 & 0.065 $\pm$ 0.01 & 0.072 & ($-11\%$ $\pm$ $19\%$) & 0.089 & ($-37\%$ $\pm$ $19\%$) & 0.079 & ($-22\%$ $\pm$ $19\%$)\\
      (total) & 220.3 & 0.75 $\pm$ 0.1 & 0.88 & ($-15\%$ $\pm$ $16\%$) & 1.25 & ($-67\%$ $\pm$ $16\%$) & 1.10 & ($-47\%$ $\pm$ $16\%$)\\
      & 300 & 1.25 $\pm$ 0.1 & 1.41 & ($-13\%$ $\pm$ $9\%$) & - & - & - & -\\
      & 2000 & 0.65 $\pm$ 0.1 & 0.75 & ($-15\%$ $\pm$ $19\%$) & - & - & - & -\\
      \hline
      $LNO$  & 110.3 & 0.016 $\pm$ 0.002 & 0.009 & ($78\%$ $\pm$ $15\%$) & 0.011 & ($45\%$ $\pm$ $15\%$) & 0.009 & ($78\%$ $\pm$ $15\%$)\\
      (total) & 220.3 & 0.3 $\pm$ 0.04 & 0.165 & ($82\%$ $\pm$ $16\%$) & 0.209 & ($44\%$ $\pm$ $16\%$) & 0.201 & ($49\%$ $\pm$ $16\%$)\\
      & 300 & 0.54 $\pm$ 0.07 & 0.30 & ($80\%$ $\pm$ $15\%$) & - & - & - & -\\
      & 2000 & 0.37 $\pm$ 0.05 & 0.21 & ($76\%$ $\pm$ $16\%$) & - & - & - & -\\
      \hline\hline
\end{longtable*}

\bibliographystyle{apjbib}

\end{document}